\documentclass[aps,prl,twocolumn,nofootinbib]{revtex4}
\usepackage{graphicx}

\newcommand{\eg}{{\it e.g.}}

\newcommand{\beq}{\begin{equation}}
\newcommand{\eeq}{\end{equation}}
\newcommand{\bea}{\begin{eqnarray}}
\newcommand{\eea}{\end{eqnarray}}
\newcommand{\nn}{\nonumber}

\begin{document}
\preprint{hep-ph/0604223}

\title{Constraints on flavor-changing $Z'$ models by $B_s$ mixing, $Z'$
  production, and $B_s \to \mu^+ \mu^-$}

\author{Kingman Cheung$^{1}$,
Cheng-Wei Chiang$^{2,3}$,
N.G. Deshpande$^{4}$,
J. Jiang$^{4}$}

\affiliation{$^1$ NCTS and Department of Physics, National Tsing Hua
  University, Hsinchu, Taiwan R.O.C.}
\affiliation{$^2$ Department of Physics, National Central University, Chungli,
  Taiwan 320, R.O.C.}
\affiliation{$^3$ Institute of Physics, Academia Sinica, Taipei, Taiwan 115,
  R.O.C.}
\affiliation{$^4$ Institute for Theoretical Science, University of Oregon,
  Eugene, OR 97403}

\date{\today}

\begin{abstract}
  Certain string-inspired $Z'$ models have non-universal interactions to three
  families of fermions and induced tree-level flavor-changing couplings.  We
  use recent results on $B_s$-$\overline{B_s}$ mixing to constrain the size of
  the flavor-changing couplings in the $b$-$s$ sector.  In some highly
  predictive $Z'$ models, such a constraint on $b$-$s$ coupling can be
  translated into the flavor-diagonal couplings.  Based on the $Z'$ production
  limits at the Tevatron, we obtain the limit on the leptonic couplings of $Z'$
  and then make predictions for $B_s \to \mu^+ \mu^-$ branching ratios.  We
  conclude that with the present constraints from $B_s$ mixing and $Z'$
  production, the muonic decay of $B_s$ may not be observed at the Tevatron if
  the projected integrated luminosity is less than $O(5-10)$ fb.
\end{abstract}

\maketitle


Searches for flavor-changing neutral currents (FCNC) have been pursued for many
years.  So far, the sizes of FCNC in the $u$-$c$, $b$-$s$, $s$-$d$, and $b$-$d$
sectors, in general, agree with the Standard Model (SM) predictions, namely,
those derived from the Cabibbo-Kobayashi-Maskawa (CKM) mechanism in higher
order.  The FCNC effect in $b$-$s$ sector was recently confirmed in the $B_s$
meson mixing observed by both CDF and D\O\ :
\begin{eqnarray}
  {\rm CDF:} && \Delta M_s = 17.33 \, ^{+0.42}_{-0.21} \,({\rm stat.}) \;
                      \pm 0.07\, ({\rm syst.}) \, {\rm ps}^{-1}\;, \nonumber \\
{\rm D\O\ :} && \Delta M_s = 19.0 \pm 1.215 \, {\rm ps}^{-1} \nonumber \;,
\end{eqnarray}
where we have converted the 90\% C.L. bound $17 < \Delta M_s < 21$ ps$^{-1}$ of
D\O\ into $1\sigma$ range assuming the error is Gaussian.  We combine both
results, again assuming Gaussian errors, and get
\begin{equation}
\Delta M_s^{\rm exp} = 17.46 \; ^{+0.47}_{-0.30} \; {\rm ps}^{-1} 
                \qquad (1 \sigma \; \mbox{range}) \;.
\end{equation}
We re-evaluate the SM prediction \cite{Barger:2004qc}, using the best-fitted
inputs (given later) before the announcement of the new $B_s$ mixing data,
\begin{equation}
\label{sm}
 \Delta M_s^{\rm SM} = 19.52  \pm 5.28 \; {\rm ps}^{-1} \;.
\end{equation}
It is important to use the best-fitted inputs without the new $B_s$ mixing data
in order to determine if there is any discrepancy between the data and the SM
prediction.  Measurement of $B_s$ mixing is often used to determine the value
of $|V_{ts}|$, but this is clearly inappropriate when the mixing has additional
contribution from new physics.  The SM prediction in Eq.~(\ref{sm}) contains
large uncertainty from the hadronic parameters, nevertheless, the data agrees
fairly well with the SM value.  Therefore, we can use the $B_s$ data to
constrain new physics that may induce the $b$-$s$ transitions.

Another important channel to search for FCNC is the muonic $B_s$ decay, $B_s
\to \mu^+ \mu^-$, which has the largest chance to be detected at hadronic
machines.  In the SM, this process is loop-suppressed.  However, many
extensions of the SM predict a branching ratio large enough to be seen at
hadron colliders.  We consider an FCNC $Z'$ model inspired by string theory in
this letter.  The $B_s$ mixing and $B_s\to \mu^+ \mu^-$ are highly correlated
because from the $B_s$ data one can constrain the FCNC $b$-$s$-$Z'$ coupling,
which is an essential element in the calculation of the muonic decay.  One
additional element is the $\mu$-$\mu$-$Z'$ coupling.  In order to make a
reliable prediction for the muonic decay branching ratio, we take into account
the $\sigma(Z') \cdot B(Z' \to e^+ e^-)$ limits from the Tevatron.  In the $Z'$
model considered here, the FCNC $b$-$s$-$Z'$ coupling is related to the
flavor-diagonal couplings $qqZ'$ in a predictive way, which are then used to
obtain the upper limits on the leptonic $\ell\ell Z'$ couplings.  Therefore, we
are able to predict the maximally allowed branching ratio for the muonic decay
of $B_s$.  The predicted branching ratio is always less than $9\times 10^{-9}$
for $M_{Z'}=200-900$ GeV.  It implies that with the present constraints from
$B_s$ mixing and $Z'$ production, the muonic decay of $B_s$ may not be
observable at the Tevatron if the projected integrated luminosity is less than
$O(5-10)$ fb.

In some string-inspired models, the three generations of SM fermions are
constructed differently and may result in family non-universal couplings to an
extra $U(1)$ gauge boson, $Z'$.  Without loss of generality, we consider the
case that the $Z'$ couples with a different strength to the third generation,
as motivated by a particular class of string models \cite{chaudhuri}.  Once we
do a unitary rotation from the interaction basis to mass eigenbasis, tree-level
FCNCs are induced naturally.  Several works have recently been done regarding
the FCNCs in the down-quark sector
\cite{Barger:2003hg,Barger:2004qc,Barger:2004hn}.  In order to increase the
predictive power, we assume that the left-handed (LH) up-type sector is already
in diagonal form, such that $V_{\rm CKM} = V_{dL}$, where $V_{dL}$ is the LH
down-type sector unitary rotation matrix.  Since we do not have much
information about either the right-handed (RH) up-type or the RH down-type
sectors, we simply assume that their interactions with $Z'$ are
family-universal and flavor-diagonal in the interaction basis.  In this case,
unitary rotations keep the RH couplings flavor-diagonal.  Therefore, the FCNCs
only arise in the LH $d$-$s$-$b$ sector.  The couplings depend on the CKM
matrix elements and one additional parameter $x$, which denotes the strength of
$Z'$ coupling to the third generation LH quarks relative to the first two
generations.  Consequently, if $x$ is an $O(1)$ parameter but not exactly equal
to 1, the $b$-$s$-$Z'$ coupling will induce a significant FCNC effect.


We follow closely the formalism given in Ref.~\cite{Langacker:2000ju}.  We
assume for simplicity that there is no mixing between $Z$ and $Z'$, as favored
by the precision data.  The current associated with the additional $U(1)$ gauge
symmetry is
\beq
  J^{(2)}_{\mu} = \sum\limits_{i,j} \overline{\psi}_i \gamma_{\mu} 
    \left[ \epsilon^{(2)}_{\psi_{L_{ij}}}P_L 
     + \epsilon^{(2)}_{\psi_{R_{ij}}}P_R\right]\psi_j\,,
\eeq
where $\epsilon^{(2)}_{\psi_{{L,R}_{ij}}}$ is the chiral coupling of $Z'$ with
fermions $i$ and $j$ running over all quarks and leptons.  The $Z'$ couplings
to the leptons and up-type quarks are assumed flavor-diagonal and
family-universal: $\epsilon_{L,R}^u = Q_{L,R}^u {\bf 1}$, $\epsilon_{L,R}^e =
Q_{L,R}^e {\bf 1}$ and $\epsilon_{L}^\nu = Q_{L}^\nu {\bf 1}$ where ${\bf 1}$
is the $3 \times 3$ identity matrix in the generation space and $Q_{L,R}^{u}$,
$Q_{L,R}^{e}$ and $Q_{L}^{\nu}$ are the chiral charges.  On the other hand, the
interaction of $Z'$ with the down-type quarks is
\begin{equation}
{\mathcal L}^{(2)}_{\rm NC} =
 - g_2  Z'_\mu  \left( \bar d, \; \bar s,\; \bar b \right)_I  \gamma^\mu
  \left( \epsilon^d_L P_L + \epsilon^d_R P_R \right )
  \left(  \begin{array}{c}
                d \\
                s \\
                b \end{array} \right )_I ,
\end{equation}
where the subscript $I$ denotes the interaction basis.  For definiteness in our
predictions, we assume
\begin{equation}
  \epsilon_L^d = Q^d_L \left( \begin{array}{ccc}
                       1 & 0 & 0 \\
                       0 & 1 & 0 \\
                       0 & 0 & x  \end{array} \right ) \; \qquad
  \epsilon_R^d = Q^d_R \left( \begin{array}{ccc}
                       1 & 0 & 0 \\
                       0 & 1 & 0 \\
                       0 & 0 & 1  \end{array} \right ) ~.   \label{epsilonLR}
\end{equation}
The deviation from family universality and thus the magnitude of FCNC are
characterized by the parameter $x$ in the $\overline{b_L}$-$b_L$-$Z'$ entry.
The chiral charges have to be specified by the $Z'$ model of interest.

When diagonalizing the down-type Yukawa matrix, we rotate the LH and RH fields
by $V_{dL}$ and $V_{dR}$, respectively.  With the form of $\epsilon^d_{L,R}$
assumed as in Eq.~(\ref{epsilonLR}), the RH sector remains flavor-diagonal in
the mass eigenbasis but the LH sector $V^\dagger_{dL}\epsilon^d_L V_{dL}$ is in
general non-diagonal.  With no mixing in up-quark sector, we have $V_{\rm CKM}
= V_{dL}$, making the model very predictive.  Explicitly,
\begin{eqnarray}
B^{d}_L
&\equiv& V^\dagger_{dL}\epsilon^d_L V_{dL}
= V^\dagger_{\rm CKM}\epsilon^d_L V_{\rm CKM}
\label{zput}   \\
&& \hspace{-0.3in} \approx  Q_L^d \left(\begin{array}{ccc}
1  & (x-1) V_{ts} V_{td}^* & (x-1) V_{tb}V_{td}^* \\
(x-1) V_{td} V_{ts}^* & 1  & (x-1) V_{tb} V_{ts}^* \\
(x-1) V_{td} V_{tb}^* & (x-1) V_{ts} V_{tb}^* & x \\
\end{array}\right) \nonumber 
\end{eqnarray}
where we have made simplifications using the unitarity of $V_{\rm CKM}$.  It is
interesting to note that the sizes of the flavor-changing couplings satisfy the
hierarchy: $|B^{bs}_L| > |B^{bd}_L| > |B^{sd}_L|$.

So far, we have not specified the RH chiral couplings of the down sector nor
the chiral couplings of the up sector.  In order to obtain constraints from
$Z'$ production at the Tevatron, we take the following assumptions
\begin{equation}
\label{assume}
|Q_{R}^d| = |Q_L^d|\,, \qquad 
|Q^u_{L,R}| = |Q^d_{L,R}| \;.
\end{equation}
Such assumptions are reasonable; many $Z'$ models predict chiral couplings to
be of a similar order ({\it e.g.}, the $Z_\psi$ model has all chiral couplings
equal to $1/\sqrt{24}$).  Moreover, the prediction of $Z'$ production depends
on the factors ${Q^{q}_L}^2 + {Q^q_R}^2$.  Thus, changing to another $Z'$ model
would not affect the limits significantly.


Within the SM the mass difference in the $B_s$ system is \cite{Buchalla:1995vs}
\begin{eqnarray}
  \Delta M_{B_s}^{\rm SM}
  &=& \frac{G_F^2}{6 \pi^2} M_W^2 m_{B_s} f_{B_s}^2 (V_{tb}V_{ts}^*)^2
    \eta_{2B} S_0(x_t) \nonumber \\
 && \hspace{-0.3in} \times [\alpha_s(m_b)]^{-6/23} 
    \left[ 1 + \frac{\alpha_s(m_b)} {4\pi} J_5 \right] B_{B_s}(m_b)
    ~,
\label{eq:m12}
\end{eqnarray}
where $S_0(x_t) = 2.463$ and the NLO short-distance QCD corrections are encoded
in the parameters $\eta_{2B} \simeq 0.551$ and $J_5 \simeq 1.627$
\cite{Buchalla:1995vs}.  We have taken $M_{B_s} = 5.3696 \pm 0.0024$ GeV, and
$\tau_{B_s} = 1.466 \pm 0.059$ ps$^{-1}$ \cite{PDG2005}.  For the Wolfenstein
parameters, we use the CKMfitter results after EPS 2005: $\lambda = 0.22622 \pm
0.00100$, and $A = 0.825^{+0.011}_{-0.019}$ , $\bar\rho =
0.207^{+0.036}_{-0.043}$, and $\bar\eta = 0.340 \pm 0.023$ \cite{CKMfitter}.
The hadronic parameter $f_{B_s} \sqrt{{\hat B}_{B_s}} = 0.262 \pm 0.035$ is
taken from the lattice calculation \cite{Hashimoto:2004hn}.  After taking the
mean for asymmetric errors, we obtain the SM prediction
\footnote{For consistency, this SM value is obtained without referring to the
  measured $\Delta M_d$ because $Z'$ can also have contributions in the $B_d$
  system, even though the uncertainty in Eq.~(\ref{sm}) could have been much
  smaller by doing so.}
as in Eq.~(\ref{sm}).  The effect of LH FCNC induced by $Z'$ is given by
\begin{equation}
\label{range}
  \frac{\Delta M_s^{\rm exp}}{\Delta M_s^{\rm SM}}
  = \left| 1 + 3.57 \times 10^5
    \left( \rho_L^{sb} \right)^2 e^{2i\phi_L^{sb}} \right|
  = 0.894 \pm 0.243 ~,
\end{equation}
where $\phi_L^{sb}$ is the weak phase associated with the coupling $B_L^{sb}$.
In the model that we consider
\begin{eqnarray}
  \rho_L^{sb} \equiv \left| \frac{g_2 M_Z}{g_1 M_{Z'}} B_L^{sb} \right|
  = \left| \frac{g_2 M_Z}{g_1 M_{Z'}} (x-1) Q_L^d V_{tb} V_{ts}^* \right| ~,
\end{eqnarray}
and $\phi_L^{sb} = 180^\circ$.  We have combined the relative errors of $\Delta
M_s^{\rm exp}$ and $\Delta M_s^{\rm SM}$ in quadrature in Eq.~(\ref{range}).

\begin{figure}[t!]
\includegraphics[width=.5\textwidth]{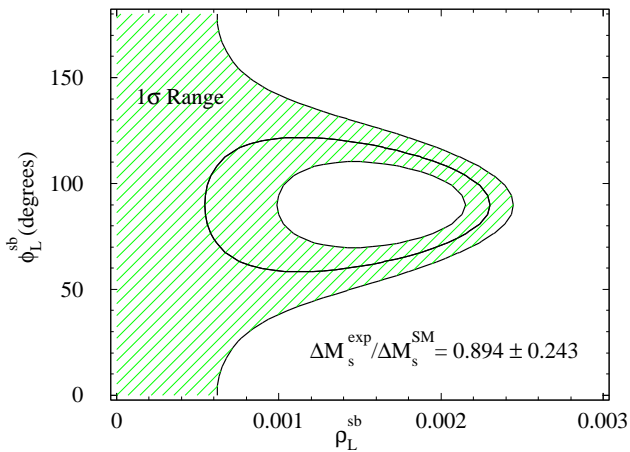}
\caption{
  The allowed parameter space in the $Z'$ model with FCNC only in the LH
  sector.  The shaded area corresponds to the $1\sigma$ C.~L.\ limits of
  $\Delta M_s^{\rm exp}/\Delta M_s^{\rm SM} = 0.894 \pm 0.243$.  The central
  value $0.894$ corresponding to $\Delta M_s^{\rm exp} = 17.46$ ps$^{-1}$ is
  also indicated.
\label{fig:contour}}
\end{figure}

We show the allowed parameter space of $(\rho_L^{sb}, \phi_L^{sb})$ in
Fig.~\ref{fig:contour}.  For $\phi_L^{sb}=0$ or $180^\circ$, $\rho_L^{sb}$ is
constrained to be less than $6.20\times 10^{-4}$.  In more general models,
$\phi_L^{sb}$ may have a different value.  For example, if $\phi_L^{sb} =
90^\circ$, $\rho_L^{sb}$ is constrained to be less than $9.87 \times 10^{-4}$.
Note that there are regions with $\rho_L^{sb} > 9.87 \times 10^{-4}$ also
allowed by the current $\Delta M_s$ constraint, \eg, $ 2.15 \times 10^{-3} \le
\rho_L^{sb} \le 2.45 \times 10^{-3}$ for $\phi_L^{sb} = 90^\circ$.  However,
some of these regions correspond to $Z'$ contributions larger than the SM
contributions.  Although not completely impossible, we think it is unlikely and
thus leave it out from the discussions in the rest of the paper.


\begin{table}[b!]
\caption{ \small \label{table2}
The 95\% C.L. limits on $\sigma(Z')\cdot B(Z'\to e^+ e^-)$ given by the
preliminary CDF result in Ref.~\cite{cdf-z} as a function of $M_{Z'}$.
}
\begin{ruledtabular}
\begin{tabular}{cccc}
$M_{Z'}$ (GeV) & $\sigma\cdot B^{95}$ (pb) & 
$M_{Z'}$ (GeV) & $ \sigma\cdot B^{95}$ (pb)\\
\hline 
200 & 0.0505 & 600 & 0.0132 \\
250 & 0.0743 & 650 & 0.0136 \\
300 & 0.0289 & 700 & 0.0134 \\
350 & 0.0404 & 750 & 0.0126 \\
400 & 0.0261 & 800 & 0.0171 \\
450 & 0.0259 & 850 & 0.0172 \\
500 & 0.0172 & 900 & 0.0215 \\
550 & 0.0138 & 950 & 0.0246 
\end{tabular}
\end{ruledtabular}
\end{table}

The production cross section of $Z'$ followed by the leptonic decay is given by
\begin{eqnarray}
\sigma(p\bar p \to Z' \to \ell^+ \ell^-) &=& 
  \frac{g_2^4}{144} \frac{1}{s} \frac{M_{Z'}}{\Gamma_{Z'}} 
 \left( {Q_L^e}^2 + {Q_R^e}^2 \right ) \nonumber \\
&&\hspace{-1.5in} \times 
\sum_{q=u,d,s,c}
 \left( {Q_L^q}^2 + {Q_R^q}^2 \right )
 \int^1_{r}\, \frac{dx}{x}  f_{q} (x) \, 
     f_{\bar q} \left( \frac{r}{x} \right )
\end{eqnarray}
where $\sqrt{s}=1960$ GeV, $r = M_{Z'}^2/s$ and $\Gamma_{Z'}$ is the total
width.  The partial width $Z' \to f \bar f$ is
\begin{eqnarray}
\Gamma (Z' \to f \bar f) &=& \frac{N_f g_2^2 M_{Z'} }{48 \pi}\, 
\sqrt{1 - 4 \mu} \, \biggr [
2 \left( |Q_L^{f}|^2 + |Q_R^{f}|^2 \right ) \nonumber \\
 &&\times (
   1 - \mu  )  + 12 \mu \, Q_L^{f} \, Q_R^{f} 
               \,  \biggr ]
\end{eqnarray}
where $N_f = 3(1)$ for quark (lepton) and $\mu = m_f^2/M_{Z'}^2$. We have
included all leptonic and hadronic modes in the total width.  Note that the
FCNC contributions are negligible and $Z'\to W^+W^-$ is highly suppressed by
the $Z$-$Z'$ mixing angle, which is severely constrained by electroweak
precision data \cite{luo}.

\begin{figure}[t]
\includegraphics[width=0.5\textwidth]{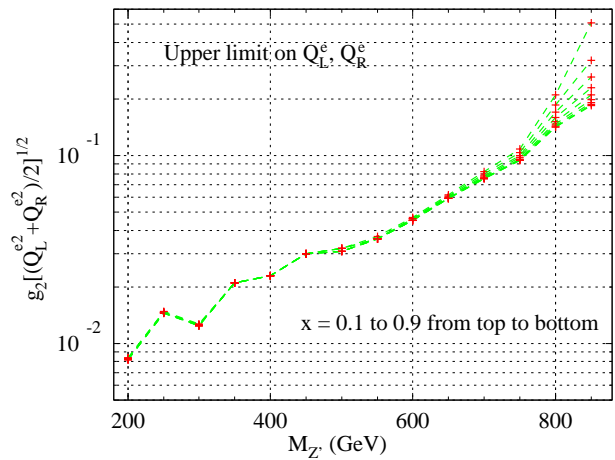}
\caption{\label{zlimit}
  The upper limits on $g_2 \sqrt{ ( {Q_L^e}^2 + {Q_R^e}^2 )/2}$ obtained using
  the CDF 95\% C.L. limits of $\sigma(Z') \cdot B(Z'\to e^+ e^-)$ tabulated in
  Table \ref{table2}.  The constrained value of $\rho_L^{sb}=6.20\times
  10^{-4}$ is used.}
\end{figure}

The most recent (though preliminary) upper limits on the $Z'$ search was
performed by CDF with an integrated luminosity $819$ pb$^{-1}$ \cite{cdf-z}.
We read off the limits of $\sigma \cdot B(Z' \to e^+ e^-)$ from their figure,
and tabulated in Table \ref{table2}.  We use the constrained value of
$\rho_L^{sb}$ to obtain the value for $g_2 Q_L^{sb}$ for each $M_{Z'}$, which
is in turned related to $g_2 Q_L^d$ by Eq.~(\ref{zput}).  With our assumptions
in Eq.~(\ref{assume}), we can then obtain the upper limits on $g_2 \sqrt{
  ({Q_L^e}^2 + {Q_R^e}^2)/2 }$ from the CDF 95\% C.L.  limits of $\sigma(Z')
\cdot B(Z'\to e^+ e^-)$.  We show the limits of $g_2 \sqrt{ ({Q_L^e}^2 +
  {Q_R^e}^2)/2 }$ with the constrained value of $\rho_L^{sb}=6.20 \times
10^{-4}$ in Fig.~\ref{zlimit}.  At smaller $M_{Z'}$ the limits are insensitive
to the value of $x$, while the difference is more visible at larger $M_{Z'}$.
When $x$ gets larger toward $1$, the allowed value for $Q_L^d$ from
$\rho_L^{sb}$ increases.  Thus, the corresponding upper limits on $Q_{L,R}^e$
have to be smaller.  That is why at each $M_{Z'}$ the chosen values of $x=0.1 -
0.9$ go from top to bottom.  The general trend for increasing $M_{Z'}$ is the
increase in the upper limit of $Q_{L,R}^e$.  This is easy to understand because
at large $M_{Z'}$ the dominant factor in the production cross section is the
parton density, which becomes very small at large momentum fractions.

\begin{figure}[t]
\includegraphics[width=.5\textwidth]{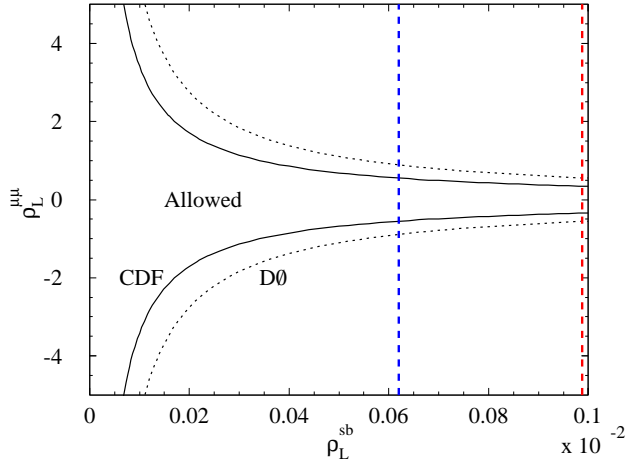}
\caption{\label{mumuspace}
  The allowed parameter space on the $(\rho_L^{sb},\rho_L^{\mu\mu}=
  \rho_R^{\mu\mu})$ plane using the current upper bounds on $B(B_s \to \mu^+
  \mu^-)$: $1.0 \times 10^{-7}$ (CDF) and $2.3 \times 10^{-7}$ (D\O).  The
  upper limit of $\rho_L^{sb}$ for $\phi^{sb}_L = 0^\circ$ and $90^\circ$ are
  also indicated using the blue dashed and red dashed lines, respectively.}
\end{figure}

We are ready to compute the maximally allowed decay rate for $B_s \to \mu^+
\mu^-$.  We ignore the RG running effect at the $b$-$s$-$Z'$ vertex, which is
good enough for an order-of-magnitude estimate.  The branching ratio of $B_s
\to \mu^+ \mu^-$ is given by
\begin{eqnarray}
B(B_s \to \mu^+ \mu^-)
&=& \tau(B_s) \frac{G_F^2}{4\pi} f_{B_s}^2 m_{\mu}^2 m_{B_s}
    \sqrt{1-\frac{4 m_{\mu}^2}{m_{B_s}^2}} |V_{tb}^* V_{ts}|^2 \nn \\
&&  \hspace{-1.3in}
\times
    \left\{
    \left| \frac{\alpha}{2\pi \sin^2\theta_W} Y \left(\frac{m_t^2}{M_W^2}\right)
      + 2 \frac{\rho_L^{bs} \rho_L^{\mu\mu}}{V_{tb}^* V_{ts}}
    \right|^2
    + \left| 2 \frac{\rho_L^{bs} \rho_R^{\mu\mu}}{V_{tb}^* V_{ts}} \right|^2
      \right\} ~,
\end{eqnarray}
where 
\begin{eqnarray}
\rho_{L,R}^{\mu\mu} = \frac{g_2 M_Z}{g_1 M_{Z'}} Q_{L,R}^e
\end{eqnarray}
at the weak scale.  One can find the definition of $Y (m_t^2 / M_W^2)$ in the
SM part in Ref.~\cite{Buchalla:1995vs}; its value is about $1.05$ here.  Using
the central value of the averaged $B_s$ lifetime $\tau_{B_s} = 1.461$ ps,
corresponding to $\Gamma_{B_s} = (4.49 \pm 0.18) \times 10^{-13}$ GeV and
$f_{B_s} = 230$ MeV, we obtain a SM branching ratio of about $4.2 \times
10^{-9}$.  The current upper limits on $B(B_s \to \mu^+ \mu^-)$ from CDF and
D\O\ based on $780$ and $700$ pb$^{-1}$ data, respectively, are \cite{D0FPCP}:
\begin{eqnarray}
B(B_s \to \mu^+ \mu^-) &<& 1.0 \times 10^{-7} \;\;\;
\mbox{(CDF)}  \nonumber \\
B(B_s \to \mu^+ \mu^-)  &<& 2.3 \times 10^{-7} \;\;\;
\mbox{(D\O\ )} ~. \nonumber
\end{eqnarray}
Given the assumption that RH FCNC couplings are ignored and the fact that
$\phi_L^{sb} = 180^\circ$ in the current model, we plot the allowed parameter
space on the $\rho_L^{sb}$-$\rho_L^{\mu\mu}$ plane in Fig.~\ref{mumuspace}.  We
have taken $\rho_R^{\mu\mu} = \rho_L^{\mu\mu}$.

\begin{figure}[t]
\includegraphics[width=.5\textwidth]{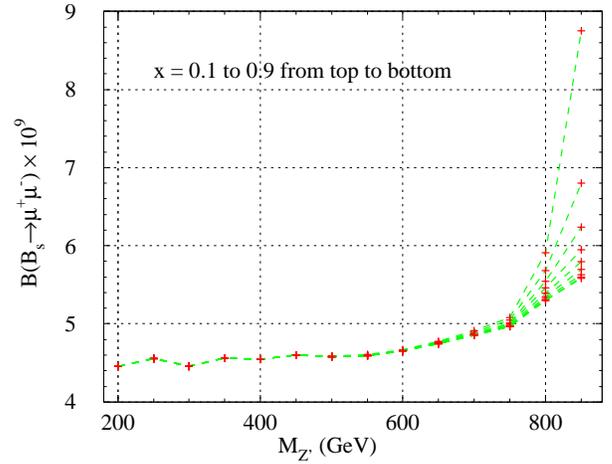}
\caption{\label{mumupred}
  Dependence of the upper limits of $B(B_s \to \mu^+ \mu^-)$ on
  $M_{Z'}$, where different sets of lines correspond to different values of $x$
  varying from $0.1$ to $0.9$. We have set 
  $\rho_R^{\mu\mu} = \rho_L^{\mu\mu}$.}
\end{figure}

If we take the upper limit of $\rho_L^{sb}=6.20 \times 10^{-4}$, and the
conservative upper limit of $g_2 Q_L^e$ given in Fig.~\ref{zlimit} for
different values of $M_{Z'}$, we obtain the upper limits of the $B_s \to \mu^+
\mu^-$ branching ratio, as shown in Fig.~\ref{mumupred}.  Different sets of
lines in the drawing correspond to different values of the parameter $x$ in our
model, ranging from $0.1$ to $0.9$.  It is seen again that there is less
variations in the predictions for small $M_{Z'}$.  The result that the
maximally allowed branching ratio is always less than $9\times 10^{-9}$ implies
that Tevatron may not have the capability to observe the muonic decay of $B_s$.
These upper limits are at least one order of magnitude below the current best
upper bound on $B(B_s \to \mu^+ \mu^-)$ quoted above.

We conclude that with the present constraints from $B_s$ mixing and $Z'$
production, the muonic decay of $B_s$ may not be observed at the Tevatron if
the projected integrated luminosity is less than $O(5-10)$ fb.  However, at
LHCb, with anticipated production of $10^{12}$ $b\bar b$ per year, the expected
branching ratio of order $10^{-9}$ is observable.

This research was supported in part by the National Science Council of Taiwan
R.O.C.\ under Grant Nos.\ NSC 94-2112-M-007-010- and NSC 94-2112-M-008-023-,
and by the National Center for Theoretical Sciences, and in part by the
U.S.~Department of Energy under Grants No DE-FG02-96ER40969.



\begin{thebibliography}{10}

\bibitem{Barger:2004qc}
  V.~Barger, C.~W.~Chiang, J.~Jiang and P.~Langacker,
  Phys.\ Lett.\ B {\bf 596}, 229 (2004).

\bibitem{chaudhuri}
  S.~Chaudhuri, S.~W.~Chung, G.~Hockney and J.~Lykken,
  Nucl.\ Phys.\ B {\bf 456}, 89 (1995);
  G.~Cleaver, M.~Cvetic, J.~R.~Espinosa, L.~L.~Everett, P.~Langacker and
  J.~Wang,
  Phys.\ Rev.\ D {\bf 59}, 055005 (1999);
  M.~Cvetic, G.~Shiu and A.~M.~Uranga,
  Phys.\ Rev.\ Lett.\  {\bf 87}, 201801 (2001);
  M.~Cvetic, P.~Langacker and G.~Shiu,
  Phys.\ Rev.\ D {\bf 66}, 066004 (2002).


\bibitem{Barger:2003hg}
  V.~Barger, C.~W.~Chiang, P.~Langacker and H.~S.~Lee,
  Phys.\ Lett.\ B {\bf 580}, 186 (2004).

\bibitem{Barger:2004hn}
  V.~Barger, C.~W.~Chiang, P.~Langacker and H.~S.~Lee,
  Phys.\ Lett.\ B {\bf 598}, 218 (2004).

\bibitem{Langacker:2000ju}
  P.~Langacker and M.~Plumacher,
  Phys.\ Rev.\ D {\bf 62}, 013006 (2000).

\bibitem{Buchalla:1995vs}
  G.~Buchalla, A.~J.~Buras and M.~E.~Lautenbacher,
  Rev.\ Mod.\ Phys.\  {\bf 68}, 1125 (1996)
  [arXiv:hep-ph/9512380].

\bibitem{PDG2005}
  S.~Eidelman {\it et al.}  [Particle Data Group],
  Phys.\ Lett.\ B {\bf 592}, 1 (2004), and updated results for 2006 Edition
  available on the web site {\tt http://pdg.lbl.gov/}.

\bibitem{CKMfitter}
  A.~Hocker, H.~Lacker, S.~Laplace and F.~Le Diberder,
  Eur.\ Phys.\ J.\ C {\bf 21}, 225 (2001) [arXiv:hep-ph/0104062].
  Updated results may be found on the web site {\tt
  http://ckmfitter.in2p3.fr/}.

\bibitem{Hashimoto:2004hn}
  S.~Hashimoto,
  Int.\ J.\ Mod.\ Phys.\ A {\bf 20}, 5133 (2005)
  [arXiv:hep-ph/0411126].

\bibitem{cdf-z}
  Information is available at 
  {\tt http://www-cdf.fnal.gov/$\sim$harper/diEleAna.html}.

\bibitem{luo}
  P.~Langacker and M.~x.~Luo,
  Phys.\ Rev.\ D {\bf 45}, 278 (1992).

\bibitem{D0FPCP}
  R.V.~Kooten, talk presented at FPCP 2006, Vancouver, Canada, April 9-12,
  2006.

\end{thebibliography}
\end{document}